\documentclass[showpacs,preprintnumbers,amsmath,amssymb,APSl,prd,nofootinbib,superscriptaddress, twocolumn]{revtex4-2}

\usepackage{dcolumn}
\usepackage{bm}
\usepackage{ifpdf}
\usepackage{hyperref}
\usepackage{bm}
\usepackage{xcolor,color,graphicx,graphics}
\usepackage[spanish,english]{babel}
\usepackage[latin1]{inputenc}
\usepackage[OT1]{fontenc}
\usepackage{latexsym,amssymb,amsmath,amsfonts}
\usepackage{makeidx}
\usepackage{epsfig,subfigure}
\usepackage{natbib}
\usepackage{epstopdf}
\usepackage{mathrsfs}
\hypersetup{colorlinks=true, linkcolor=blue, citecolor=green}
\usepackage{enumerate}
\usepackage{xcolor}
 \usepackage{multirow}

\definecolor{red}{rgb}{1,0,0}

\def\+{^\dagger}

\def\<{\leftarrow}
\def\>{\rightarrow}

\def\({\left(}
\def\){\right)}

\def\arcsinh{\mathop{\rm arcsinh}\nolimits}

\newcommand{\bi}{\begin{itemize}} 				\newcommand{\ei}{\end{itemize}}
\newcommand{\benu}{\begin{enumerate}} 		\newcommand{\enu}{\end{enumerate}}
\newcommand{\bd}{\begin{dinglist}{0}}     \newcommand{\ed}{\end{dinglist}}
\newcommand{\bfig}{\begin{figure}[htbp]}  \newcommand{\efig}{\end{figure}}
        			
\newcommand{\bc}{\begin{center}} 				  \newcommand{\ec}{\end{center}}
\newcommand{\be}{\begin{equation}} 				\newcommand{\ee}{\end{equation}}
\newcommand{\bsub}{\begin{subequations}}  \newcommand{\esub}{\end{subequations}}
\newcommand{\ben}{\begin{eqnarray}} 			\newcommand{\een}{\end{eqnarray}}
\newcommand{\ba}[1]{\begin{array}{#1}} 		\newcommand{\ea}{\end{array}}
\newcommand{\bea}{\begin{equation}\begin{array}{rcl}}
\newcommand{\eea}{\end{array}\end{equation}}

\begin{document}
\title{Geodesic completeness of effective null geodesics in regular space-times with non-linear electrodynamics}

\author{Merce Guerrero} \email{merguerr@ucm.es}
\affiliation{Departamento de F\'isica Te\'orica and IPARCOS,
	Universidad Complutense de Madrid, E-28040 Madrid, Spain}
	\author{Gonzalo J. Olmo} \email{gonzalo.olmo@uv.es}
\affiliation{Departamento de F\'{i}sica Te\'{o}rica and IFIC, Centro Mixto Universidad de Valencia - CSIC.
Universidad de Valencia, Burjassot-46100, Valencia, Spain}
\affiliation{Universidade Federal do Cear\'a (UFC), Departamento de F\'isica,\\ Campus do Pici, Fortaleza - CE, C.P. 6030, 60455-760 - Brazil.}

\author{Diego Rubiera-Garcia} \email{drubiera@ucm.es}
\affiliation{Departamento de F\'isica Te\'orica and IPARCOS,
	Universidad Complutense de Madrid, E-28040 Madrid, Spain}

\date{\today}
\begin{abstract}
We study the completeness of light trajectories in certain spherically symmetric regular geometries found in Palatini theories of gravity threaded by non-linear (electromagnetic) fields, which makes their propagation to happen along geodesics of an effective metric. Two types of geodesic restoration mechanisms are employed: by pushing the focal point to infinite affine distance, thus unreachable in finite time by any sets of geodesics, or by the presence of a defocusing surface associated to the development of a wormhole throat. We discuss several examples of such geometries to conclude the completeness of all such effective paths. Our results are of interest both for the finding of singularity-free solutions and for the analysis of their optical appearances e.g. in shadow observations.

\end{abstract}

\maketitle

\section{Introduction}

Together with the information loss problem resulting from the evaporation of black holes via Hawking radiation \cite{Hawking:1975vcx}, the issue of space-time singularities, understood as the incompleteness of geodesic trajectories \cite{WaldBook}, is arguably one of the most important theoretical aspects of Einstein's General Theory of Relativity (GR) yet to be understood. Far from being mathematical artifices or the consequence of excessively idealized settings, such singularities were proved in a number of theorems to be unavoidable and persistent features of the theory \cite{Senovilla:2014gza}. Indeed, they arise in some of its most physically appealing solutions: the interior of black holes and the early Universe. Being GR a classical theory, the presence of such singularities entails its own demise via the lack of predictability and determinism they bear within. Despite the efforts of the theoreticians for more than five decades, this problem stubbornly refuses to be cracked open.

In order to restore geodesic completeness, some mechanisms have been introduced in the literature to prevent the development of a focal point, roughly split into two classes (see \cite{Carballo-Rubio:2019fnb} for an overall discussion). In the first class, the focal point is displaced to infinite affine distance, thus rendering it unreachable to any set of observers and depriving the potentially pathological region of any physical meaning. In the second class, the singular region evolves into a surface of non-vanishing area placed at a finite affine distance. This allows null (photons) and time-like (free-falling observers) trajectories to pass through it and explore other regions of the geometry. This mechanism is typically ascribed to wormholes \cite{VisserBook}. An alternative viewpoint is to blame divergences in the curvature scalars as the responsible for the development of a singularity. Thus, such a procedure replaces the ill-behaved regions of the geometries by curvature-smooth ones like de Sitter cores \cite{Ansoldi:2008jw}. Examples of this are the celebrated Bardeen \cite{Bardeen} and Hayward \cite{Hayward:2005gi} solutions (see e.g. \cite{Lemos:2011dq} for a more elaborated discussion). Either way, the hypotheses of the theorems still apply, so typically one requires a violation of the energy conditions to remove singularities unless we move to gravitational theories beyond GR.

This work is framed within the geodesic restoration achieved in certain families of spherically symmetric geometries supported by electromagnetic fields. Such families cover all the sub-cases of singularity-removal discussed above, and arise in the Palatini formulation of gravity, where metric and affine connection are a priori independent entities. Implementations of the first mechanism naturally emerge in theories of $f(R)$-type \cite{Bejarano:2017fgz}, while the second can be found in theories with additional contributions in the Ricci tensor, such as Eddington-inspired Born-Infeld (EiBI) \cite{Olmo:2015bya} gravity. While in the former one needs not to worry about what the curvature scalars are doing at the boundary of the space-time (since it is unreachable in finite affine time), this is not so in the latter since one must guarantee every time-like (physical) observer to survive its encounter with the wormhole throat to pass to the other region of the space-time. This way one is led to consider the harmlessness of  tidal forces \cite{Olmo:2016fuc}, the well-posedness of the scattering of waves, or the completeness of observers with arbitrary (but bound) accelerations \cite{Olmo:2017fbc} in these geometries as complementary regularity criteria.

The main goal of the present paper is to complete the above regularity analysis by introducing yet another feature of these geometries related to the fact that some of them are threaded by non-linear electrodynamics (NED) fields. In such a case light rays behave as if they were propagating through a dispersive medium \cite{Dittrich:1998fy, Shore:1995fz}, where the non-linear effects turn such a medium into an effective geometry \cite{Novello:1999pg,DeLorenci:2000yh}. This fact has important phenomenological consequences whenever non-linear fields are present, such as in astrophysical and cosmological environments \cite{Obukhov:2002xa,Marklund:2006my,Agostini:2011qk,Ovgun:2017iwg,Allahyari:2019jqz,Schee:2019gki,Stuchlik:2019uvf,Hu:2020usx,Breton:2021mju,Ishlah:2023tzt}. For the sake of this work, consistency of the singularity-removal mechanisms introduced above demands studying the completeness of these effective geodesics under the presence of NED fields. We shall consider examples of the two mechanisms for various choices of the gravity and matter sectors and show that geodesic restoration is also met in the effective metric. The results of our analysis will wrap up our analysis of regular black hole geometries in these theories, while at the same time prepare the ground for their astrophysical applications.

\section{Null geodesic motion with non-linear electromagnetic fields}

This work is framed on NED models defined by Lagrangian densities of the form
\begin{equation}
\varphi=\varphi(X,Y) \ ,
\end{equation}
where $\varphi(X,Y)$ is a sufficiently smooth function of the two field invariants that can be constructed in four space-time dimensions out of the field strength tensor, $F_{\mu\nu}=\partial_\mu A_{\nu}-\partial_{\nu}A_{\mu}$, and its dual, $F^{\star \mu\nu}=\tfrac{1}{2}\epsilon^{\mu\nu\alpha\beta}F_{\alpha\beta}$, as
\begin{equation} \label{eq:fieldinv}
X=-\frac{1}{2}F_{\mu\nu}F^{\mu\nu} \quad ; \quad Y=-\frac{1}{2}F_{\mu\nu}F^{\star \mu\nu} \ .
\end{equation}
Since in this work we are interested in purely electric configurations, where only $F^{tr} \neq 0$, for which $Y=0$, we neglect this invariant from now on. For these configurations, the field equations read
\begin{equation}
\nabla_{\mu}(\varphi_X F^{\mu\nu})=0 \ ,
\end{equation}
where $\varphi_X \equiv d\varphi/dX$. For any spherically symmetric geometry (the case of interest in this work), the above equation can be solved via a quadrature as
\begin{equation} \label{eq:Xr}
X\varphi_X^2=\frac{Q^2}{r^4} \ ,
\end{equation}
where $Q$ is an integration constant identified as the electric charge associated to a given configuration.

From the requirement of continuity of the electromagnetic field (but not necessarily of its first derivative) across a given surface of discontinuity, the results of \cite{Novello:1999pg,DeLorenci:2000yh} consistently probe that under this framework light rays propagate on an effective metric as
\begin{equation} \label{eq:proeq}
g_{\mu\nu}^{eff} k^\mu k^\nu=0 \ ,
\end{equation}
where $k^\mu$ is the photon's wave vector and
\begin{equation} \label{eq:gmunuef}
g_{\mu\nu}^{eff}=\varphi_X g_{\mu\nu} + 2 \varphi_{XX}F_{\mu\alpha}{F^\alpha}_{\nu} \ ,
\end{equation}
is the effective metric with $g_{\mu\nu}$ being the background geometry. For non-trivial NED functions, the non-conformal structure of this equation implies a different causal cone of photons as compared to usual Maxwell electrodynamics. Let us conveniently parameterize our spherically symmetric geometry  as
\begin{equation} \label{eq:geomnull}
ds_b^2=-A(x)dt^2+B(x)dx^2 +r^2(x)d\Omega^2 \ ,
\end{equation}
where the functions $A(x)$, $B(x)$ and $r^2(x)$ characterize the background geometry. Though spherical symmetry allows the line element to be written in terms of two independent functions only, for the moment we shall stick to Eq.(\ref{eq:geomnull}), in order to accommodate a larger flexibility in the way the  metric functions are written, in particular, a non-trivial radial function $r^2(x)$. The effective version of the background geometry (\ref{eq:geomnull}) is parameterized as
\begin{equation} \label{eq:geomnulle}
ds_{eff}^2=H(x)(-A(x)dt^2+B(x)dx^2) +h(x)r^2(x)d\Omega^2 \ ,
\end{equation}
where $H(x)$ and $h(x)$ are functions determined by the (effective) null geodesic metric \eqref{eq:gmunuef} as
\begin{equation}\label{eq: H def}
H(x)=\varphi_X +2X\varphi_{XX} \quad ; \quad h(x)=\varphi_X \ .
\end{equation}
Assuming motion in the plane $\theta=\pi/2$ without loss of generality given the spherical symmetry of the space-time,  the propagation equation (\ref{eq:proeq}) takes the form
\begin{equation} \label{eq:geo1}
-AH\dot{t}^2+BH\dot{x}^2+hr^2\dot{\phi}^2=0 \ ,
\end{equation}
where a dot denotes derivative with respect to the affine parameter $u$. There are two conserved quantities: the generalized momenta associated to the $t$ and $\theta$ coordinates of the corresponding Hamiltonian, and read
\begin{equation}
E=HA\dot{t} \quad ; \quad L=hr^2\dot{\phi} \ ,
\end{equation}
interpreted as the energy and angular momentum per unit mass, respectively. Using these quantities, the geodesic equation (\ref{eq:geo1}) is written as
\begin{equation} \label{eq:geomain}
ABH^2\left(\frac{dx}{du}\right)^2=E^2-\frac{L^2AH}{hr^2} \ .
\end{equation}
This is the main equation we need in order to study the restoration of geodesic completeness in the effective metric, which we tackle in the next section.

\section{Effective geodesic completeness in $f(R)$ gravity}

We shall first consider examples of the implementation of the first mechanism for the restoration of geodesic completeness within Palatini gravities, which comes out naturally within the $f(R)$ family. In every Palatini theory belonging to the so-called Ricci-based class \cite{Afonso:2018bpv},  the affine connection can be removed in favor of the metric and the matter fields such that the new gravitational dynamics is engendered by additional terms that depend non-linearly on the energy-momentum tensor; in the $f(R)$ case only via its trace as $R=R(T)$. Since Maxwell's Lagrangian has $T=0$, this naturally pushes one to employ NEDs and, in turn, it justifies the analysis of effective geodesics. For the sake of this section, we shall base our description of the background geometry supported by Born-Infeld \cite{Bambi:2015zch} and Euler-Heisenberg \cite{Guerrero:2020uhn} NEDs, both satisfying standard energy conditions.

The corresponding background geometry can be cast in a convenient way by introducing a dimensionless radial function $z=r/r_c$, where $r_c$ contains the main parameters of the model (mass, charge, and theories' parameters). The functions (\ref{eq:geomnull}) in such a case take the form
\begin{equation} \label{eq:backfR}
A(z)=\frac{D(z)}{f_R(z)} ; \hspace{0.10cm} B(z)=\frac{1}{D(z)f_R(z)}; \hspace{0.10cm} z^2(x)=\frac{x^2}{f_R(z)} \ ,
\end{equation}
where $D(z)$ is given by an expression of the form
\begin{equation} \label{eq:DZf}
D(z)=1-\frac{1+\delta_1 G(z)}{\delta_2 z f_{R}^{1/2}(z)} \ ,
\end{equation}
with $ \{ \delta_1, \delta_2 \}$ constants and $ G(z) $ depending on the choice of the gravitational plus matter model.  In order to write the geodesic equation (\ref{eq:geomain}) in a suitable way to our purposes, we first take a derivative in the radial function defined in (\ref{eq:backfR}) to find
\begin{equation}
\frac{dx}{dz}=f_{R}^{1/2} \Big(1+\frac{zf_{R,z}}{2f_R} \Big) \ .
\end{equation}
If we apply it upon (\ref{eq:geomain}) we arrive at
\begin{equation} \label{eq:geoefff}
\frac{d\tilde{u}}{dz}=\frac{f_R^{1/2}\Big(1+\frac{zf_{R,z}}{2f_R}\Big) H}{\sqrt{E^2f_R^2-Df_R\frac{L^2}{r_c^2z^2}\frac{H}{h}}} \ ,
\end{equation}
where $\tilde{u} \equiv u/r_c$. This is the natural generalization of the original (Palatini) $f(R)$ geodesic equation, recovered when $H=h=1$ from its effective version, while the usual GR one is recovered in the limit $f_R \to 1$.

As we shall see with explicit examples, in order to restore geodesic completeness one needs the radial function $z(x)$ defined in Eq.(\ref{eq:backfR}) to have a minimum $z=z_c$ for a certain $x$ (which can be taken as $x=0$ without any loss of generality). From Eq.(\ref{eq:backfR}), this is translated into the function $f_R$ itself having a zero. In the simplest case of a quadratic gravity theory of the form
\begin{equation}\label{Eq. fR}
f(R)=R -\gamma R^2 \ ,
\end{equation}
where $\gamma$ is a quadratic-length scale, the trace equation provides $R=-\kappa^2 T$ (same as in GR), so that one finds
\begin{equation} \label{eq:f_R}
f_R=1-2\gamma R= 1 + 2\kappa^2 \gamma T \ ,
\end{equation}
where for NEDs
\begin{equation}
T=\frac{1}{2\pi}[\varphi-X\varphi_X] \ ,
\end{equation}
which is negative for both NEDs considered here. This means that such a zero could be present in the branch $\gamma>0$, so  we shall stick to that choice from now on.

\subsection{$f(R)$ gravity with Born-Infeld electrodynamics}

\begin{figure*}[t!]
\begin{center}
\includegraphics[width=8.8cm,height=5cm]{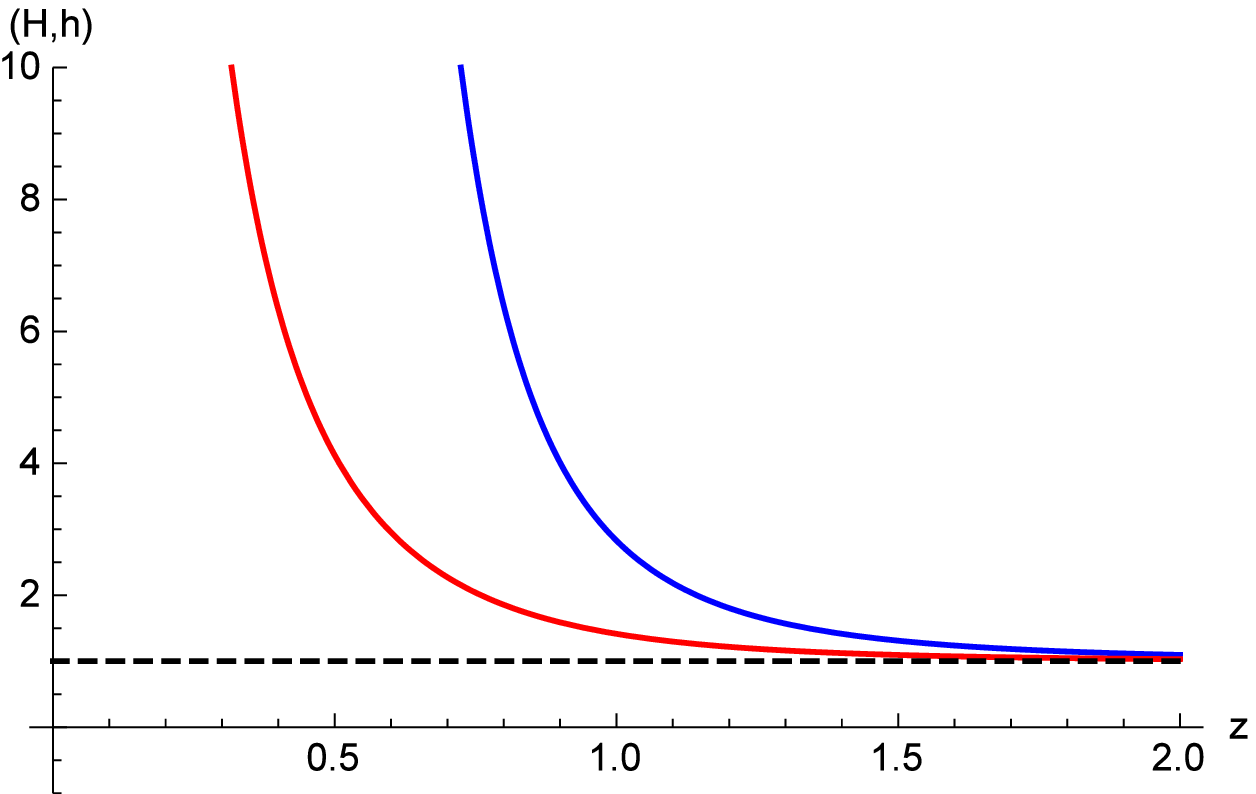}
\includegraphics[width=8.8cm,height=5cm]{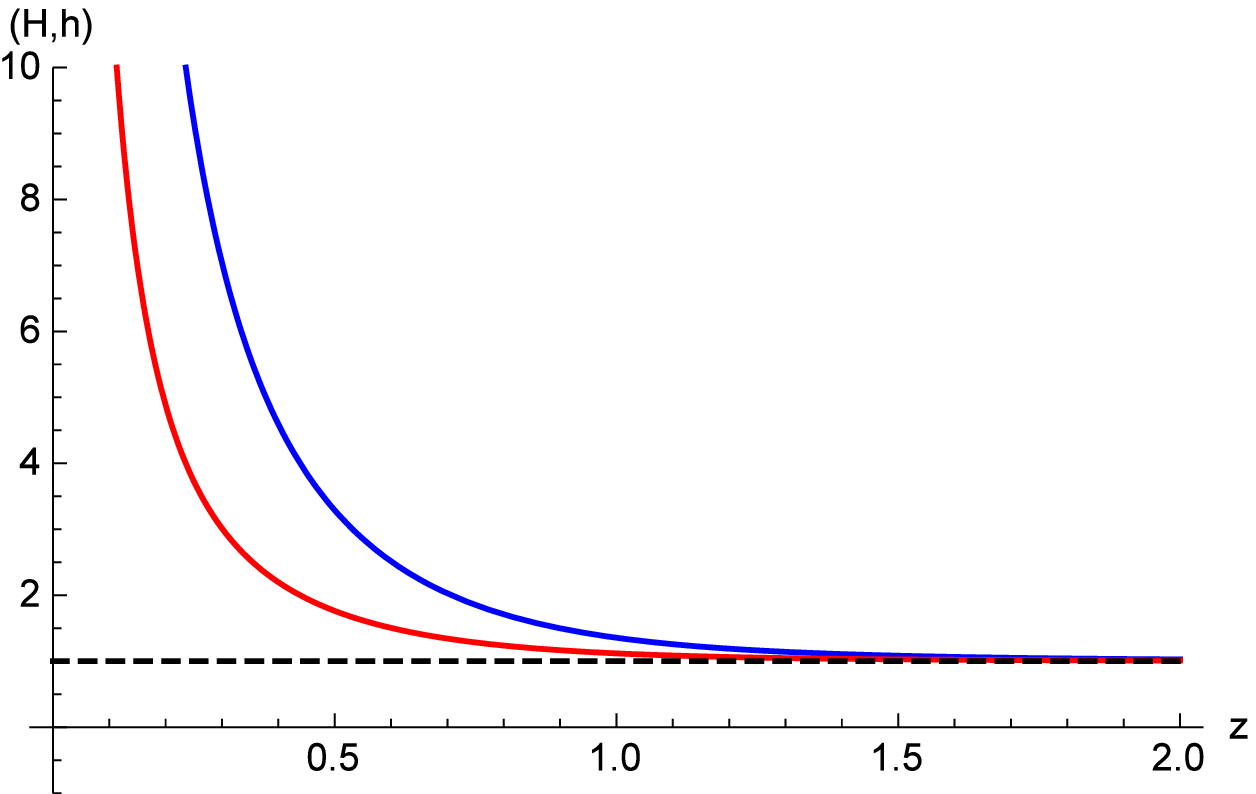}
\caption{The functions $H(z)$ (red) and $h(z)$ (blue) for BI (left) and EH (right) NEDs. Both functions are finite for every $z>0$ and bounded from below by unity, attained at  $z \to \infty$. This means that $H(z_c)$ and $h(z_c)$ are always finite and positive. }
\label{fig:Hh}
\end{center}
\end{figure*}

The Born-Infeld (BI) model is given by the function
\begin{equation}
\varphi(X)=2\beta^2 \left(1-\sqrt{1-\frac{X}{\beta^2}}\right) \ .
\end{equation}
Using Eq.(\ref{eq:Xr}) one finds that for this theory
\begin{equation}
X=\frac{\beta^2 Q^2}{\beta^2 r^4+Q^2}\ .
\end{equation}
Let us introduce dimensionless variables as $r_c^4= (4\pi) r_Q^2 l_{\beta}^2$ with $r_Q^2=\kappa^2 Q^2/(4\pi)$ the charge radius and $l_{\beta}^2=1/(\kappa^2 \beta^2)$ the BI length. In this notation, the field invariant reads $X=\tfrac{\beta^2}{1+z^4}$ while the zeros of $f_R$ in Eq.(\ref{eq:f_R}) are found at the locations (see \cite{Bambi:2015zch} for details)
\begin{equation} \label{eq:zcBI}
z_c(\lambda)=\frac{1+2\lambda -\sqrt{1+4\lambda}}{2\sqrt{1+4\lambda}} \ ,
\end{equation}
where the constant $\lambda \equiv \tfrac{\gamma}{2\pi l_{\beta}^2}$ encodes contributions from the gravity and matter sectors. From these equations one can expand the function $f_R$ in (\ref{eq:f_R}) around $z=z_c$ as
\begin{equation} \label{eq:fRexp}
f_R \approx c(\lambda)(z-z_c) + \mathcal{O}(z-z_c)^2\ ,
\end{equation}
where $c(\lambda)>0$ is a constant with an involved dependence on $\lambda$, but only its sign is relevant here. Similarly, the expansion of the function $D(z)$ at $z=z_c$ reads
\begin{equation} \label{eq:DRexp}
D(z) \approx \frac{\delta_1 a(\lambda)}{\delta_2 z_c c(\lambda)(z-z_c)^2}\ ,
\end{equation}
with $a(\lambda)>0$ is another constant in the expansion, while $\delta_1,\delta_2$ are those appearing in (\ref{eq:DZf}) and read explicitly here:
\begin{equation}
\delta_1=2(4\pi)^{3/4} \frac{r_Q}{r_S} \sqrt{\frac{r_Q}{l_{\beta}}} \quad ; \quad \delta_{2}=\frac{r_c}{r_S} \ .
\end{equation}
The final ingredient in our analysis are the expressions of the functions $h$ and $H$ defined in (\ref{eq: H def}), which read here as
\begin{equation} \label{eq:hBIzc}
h(z)=\frac{(1+z^4)^{1/2}}{z^2} \hspace{0.1cm} ; \hspace{0.1cm}
H(z)= \frac{(1+z^4)^{3/2}}{z^6} =h(z)^3 \ ,
\end{equation}
and whose behaviours are depicted in Fig.\ref{fig:Hh} [left].

\begin{figure}[t!]
\begin{center}
\includegraphics[width=8.5cm,height=5.5cm]{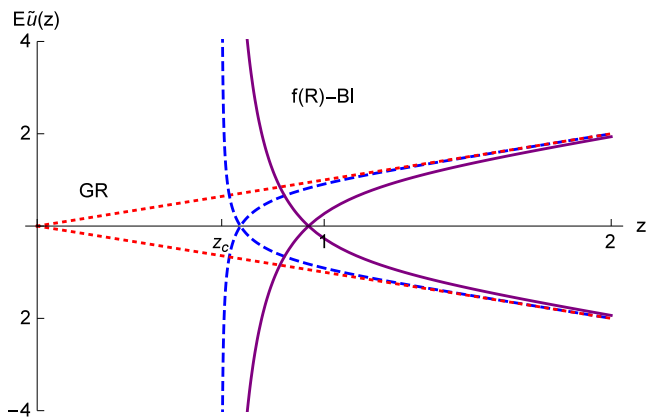}
\caption{Radial null effective geodesics (ingoing and outgoing)  for BI NED coupled to quadratic $f(R)$ gravity taking $\lambda=1$ (solid purple), as compared to background null trajectories (dashed blue, studied in \cite{Bambi:2015zch}). In red, we  depict the GR behaviour.}
\label{fig:nullradialBI}
\end{center}
\end{figure}

We have now everything ready to check the null effective geodesic completeness. Let us start with the radial case, $L=0$ in Eq.(\ref{eq:geoefff}). The result of the numerical integration is depicted in Fig. \ref{fig:nullradialBI} taking $\lambda=1$ for the effective null geodesics (solid purple) curve. As a comparison, we also depict the corresponding curves for the geodesics in the background geometry (dashed blue), described in Ref.\cite{Bambi:2015zch}. For large values of the radial function, $z \to \infty$, both types of curves converge to the GR result (dotted red), $d\tilde{u}/dz \approx \pm 1/E \to \tilde{u}(z) \approx z/E$, but they strongly diverge from it as the region $z=z_c$ is approached. Actually, in this region we can make use of the series expansions (\ref{eq:fRexp}) to find the behaviour of the geodesic equation (with $L=0$) as
\begin{equation} \label{eq:nullradialfR}
E\frac{d\tilde{u}}{dz} \simeq \pm \frac{z_c H(z_c)}{2c^{1/2} (z-z_c)^{3/2}} \ ,
\end{equation}
which is trivially integrated as
\begin{equation}
E\tilde{u} \simeq \mp \frac{z_c H(z_c)}{c^{1/2}(z-z_c)^{1/2}} \ .
\end{equation}

From this behaviour and the finiteness of $H(z)$ at every (physically accessible) $z \geq z_c$ [see Fig. \ref{fig:Hh}, left], it is clear that in this region the affine parameter grows to $u \to \pm \infty$; in other words, it can be indefinitely extended into the future (ingoing) or into the past (outgoing), which entails its completeness. Physically we interpret this as the surface $z=z_c$ being the actual (infinitely-displaced) boundary of the manifold, since it cannot be reached by any such geodesics in finite affine time. On the contrary, in GR every such geodesic gets in finite time to $z=0$ and meets its end there, thus entailing its incompleteness.

To verify the last statement for non-radial geodesics, we go back to Eq.(\ref{eq:geoefff}) and consider trajectories with $L \neq 0$. In such a case, around the region $z=z_c$ it behaves as
\begin{equation} \label{eq:LfR}
\frac{d\tilde{u}}{dz} \approx \pm \frac{c^{1/2}z_c H(z_c)}{ 2\sqrt{c^2E^2(z-z_c)^3-\frac{\delta_1 a L^2}{\delta_2 r_c^2 z_c^3 }\frac{H(z_c)}{h(z_c)}}} \ .
\end{equation}
Given the fact that both $h(z_c)$ and $H(z_c)$ are finite and positive, as follows by evaluating Eq.(\ref{eq:hBIzc}) at the boundary (\ref{eq:zcBI}), this means that the second term under the square-root is negative and, since the first one vanishes as $z \to z_c$, the square-root will always vanish at some $z$ larger than $z_c$. Thus, at some $z>z_c$ a turning point is reached, and the particle is deflected back to asymptotic infinity, no matter what its energy $E$ and angular momentum $L$ may be.

\subsection{$f(R)$ gravity with Euler-Heisenberg electrodynamics}

Let us now move to the Euler-Heisenberg (EH) electrodynamics case, described by the function ($\beta>0$)
\begin{equation}\label{eq: EH electro}
\varphi(X)= X + \beta X^2 \ .
\end{equation}
Introducing the dimensionless variables in this case as $r_c^4=54\pi l_{\beta}^2 r_Q^2$ with $l_{\beta}^2=\beta/\kappa^2$, the solution of the field equations \eqref{eq:Xr} can be expressed as (see \cite{Guerrero:2020uhn} for details)
\begin{eqnarray}
	X(z)&=&\frac{2}{3\beta} \tau^2(z) \ , \\
	\tau(z) &=& \text{Sinh}\Big[\frac{1}{3} \ln \Big[\frac{1}{z^2}\left(1+ \sqrt{z^4+1}\Big)\Big] \right] \label{eq:tau}, 
\end{eqnarray}
while the $f_R$ function is given by
\begin{equation}
f_R=1-\tilde{\gamma} \tau^4\ ,
\end{equation}
where $\tilde{\gamma}= \tfrac{4\gamma}{9\pi l_{\beta}^2}$. Its zeros are located at
\begin{equation} \label{eq:zcEH}
z_c(a)=\sqrt{\frac{2a}{a^2-1}}\ ,
\end{equation}
where the constant
\begin{equation}
a=\exp[3\arcsinh( \tilde{\gamma}^{-1/4})]\ .
\end{equation}
The expansion of the relevant functions at this location are as follows: $f_R(z)$ and  $D(z)$ read as in Eq.(\ref{eq:fRexp}) and (\ref{eq:DRexp}), respectively, with
\begin{equation}
c(\lambda)=\frac{8 \coth [\tau(z_c)]}{3z_c \sqrt{z_c^4+1}} \quad;\quad
\delta_1 \equiv \frac{(54\pi)^{3/4}}{2r_S} \sqrt{\frac{r_Q^3}{l_{\beta}}}  \ ,
\end{equation}
while $ \delta_{2} $ has the same definition as before. The final ingredients are the factors $H$ and $h$, which read as
\begin{eqnarray}
H(z)&=&2 \cosh \Bigg(\frac{2}{3} \log \Bigg(\frac{\sqrt{z^4+1}+1}{z^2}\Bigg)\Bigg)-1  \ , \\
h(z)&=& \dfrac{4}{3}\cosh \left(\frac{2}{3} \log \left(\frac{\sqrt{z^4+1}+1}{z^2}\right)\right)+1
\end{eqnarray}
being again everywhere positive and finite, in particular, at every $z=z_c(a)$, as it is depicted in Fig. \ref{fig:Hh} [right].

\begin{figure}[t!]
\begin{center}
\includegraphics[width=8.5cm,height=5.5cm]{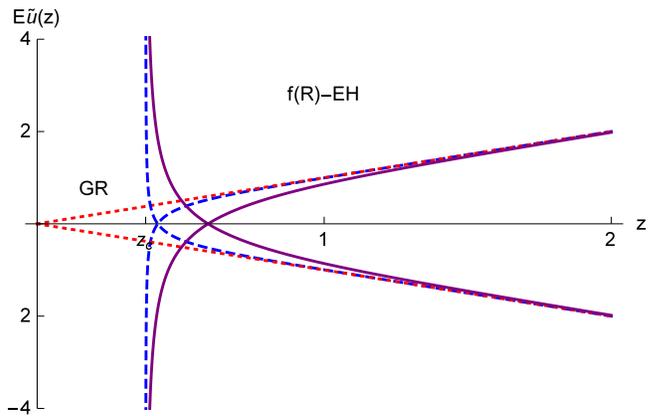}
\caption{Radial null effective geodesics (ingoing and outgoing)  for EH NED coupled to quadratic $f(R)$ gravity taking $\tilde{\gamma}=1$ (solid purple), as compared to background null trajectories (dashed blue, studied in \cite{Guerrero:2020uhn}). In red, we depict the GR behaviour.}
\label{fig:nullradialEH}
\end{center}
\end{figure}

Since the functions $f_R(z)$ and $D(z)$ have the same functional dependence as their BI counterparts, and the finiteness and positiveness of the functions $H(z)$ and $h(z)$ hold everywhere, the radial null geodesic equation (\ref{eq:nullradialfR}) is also formally the same. This way, we expect these geodesics to display a similar behaviour, something confirmed by its numerical integration depicted in Fig. \ref{fig:nullradialEH} for (ingoing and outgoing) effective null geodesics, and compared with the background geodesics, and the GR case. Therefore, our comments regarding their completeness in the BI case by displacement of the surface $z=z_c$ to the  infinitely-far away boundary of the manifold also hold here. For $L \neq 0$, effective geodesics also satisfy Eq.(\ref{eq:LfR}), which means that a turning point $z(a)>z_c(a)$ will also be attained by any geodesics, preventing them from reaching the region $z=z_c$ in pretty much the same way as in the BI case above.

\section{Effective geodesic completeness in Born-Infeld gravity}

\begin{figure}[t!]
\begin{center}
\includegraphics[width=8.5cm,height=5.8cm]{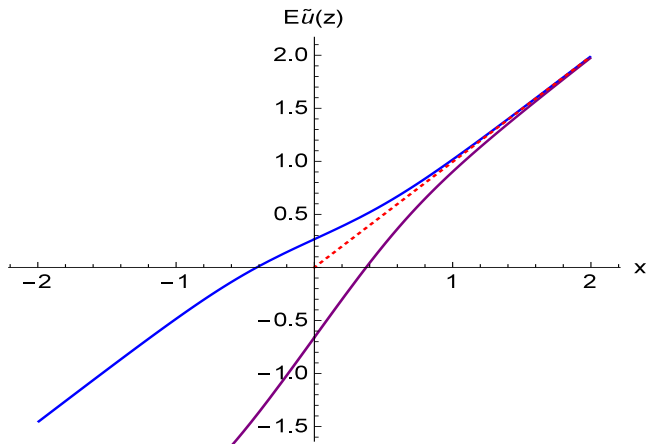}
\caption{Radial null effective geodesics (outgoing)  for EH NED coupled to EiBI gravity taking $l_{\epsilon}^2=-1$ (solid purple), as compared to background null trajectories (solid blue, studied in \cite{Guerrero:2020uhn}). In red, we  depict the GR behaviour.}
\label{fig:nullradialEHEiBI}
\end{center}
\end{figure}

We now consider an implementation of the second mechanism, in which the surface $z=z_c$ can be reached in finite affine time by null (background and effective) geodesics. This is typically the case of Palatini theories having further contributions in the Ricci tensor and, despite the fact that in such a case one finds modified gravitational dynamics even for traceless matter sources, for the sake of comparison with the $f(R)$ case we shall consider a specific setting of EiBI gravity with action (a detailed account of the properties of this theory can be found in \cite{BeltranJimenez:2017doy})
\begin{equation}
\mathcal{S}_{EiBI}=\frac{1}{\kappa^2 \epsilon} \int d^4 x \left(\sqrt{ - \text{det} (g_{\mu\nu} + \epsilon R_{\mu\nu}) } - \sqrt{-g} \right)
\end{equation}
coupled to EH electrodynamics. The completeness of background null geodesics for this combination was proven in \cite{Guerrero:2020uhn}. In this case, the geometrical functions are given by
\begin{eqnarray}
A&=&\frac{D}{\Omega_+} \hspace{0.1cm} ; \hspace{0.1cm} B=\frac{1}{\Omega_+ D} \hspace{0.1cm};\hspace{0.1cm}
D=1-\frac{1+\delta_1 G(z)}{\delta_2 z \Omega_{-}^{1/2}}  \ ,\\
\Omega_{+}&=&1-l_\epsilon^2 \tau^2(1+2\tau^2/3) \ ,\\
\Omega_{-}&=&1+l_\epsilon^2 (1+2\tau^2) \ ,
\end{eqnarray}
where $ l_\epsilon^2 = \tfrac{\epsilon}{12\pi l^2_\beta} $, while $ l^2_\beta $ has the same definition as in the previous section. This time, the transformation of radial coordinates reads
\begin{equation}
x^2=z^2 \Omega_{-} \ ,
\end{equation}
and a minimum in the radial function $z(x)$ is found whenever $\Omega_-$ develops a zero. This turns out to be the same $z=z_c$ as in Eq.(\ref{eq:zcEH}) but now with [note that this only happens in the branch  $l_{\epsilon}^2<0$]
\begin{equation}
a=\exp \Big[3\arcsinh \Big(\tfrac{1}{2} \Big(\tfrac{ (\vert l_{\epsilon}^2 \vert+8 }{\vert l_{\epsilon}^2 \vert} \Big)^{1/2}-1 \Big)^{1/2} \Big] \ .
\end{equation}
Playing the same manipulations as in the $f(R)$ case, but with the above definitions, one arrives at the effective geodesic equation in this case as
\begin{equation} \label{eq:uzEiBI}
\frac{d\tilde{u}}{dz}=\pm \frac{\Omega_{-}^{1/2} \left(1+\frac{z\Omega_{-,z}}{2\Omega_{-}}\right)H(z)}{\Omega_{+} \sqrt{E^2-\frac{DL^2}{\Omega_{+} r_c^2 z^2}\frac{H(z)}{h(z)}}} \ .
\end{equation}
In the radial case, $L=0$, this equation simplifies down to something that we can integrate near $z=z_c$; using the expansions in that limit $\Omega_{+} \approx \omega_{+}(z_c) +  \mathcal{O}(z-z_c)$ and $\Omega_{-} \approx \omega_- (z-z_c)+ \mathcal{O}(z-z_c)^2$ (with $\omega_{\pm}(z_c)$ some constants, see \cite{Guerrero:2020uhn}), this leads to the integrated equation
\begin{equation}
\pm E(\tilde{u}-\tilde{u}_0) \approx \frac{\omega_{-}^{1/2} z_c H(z_c)}{\omega_+} \sqrt{z-z_c} \ .
\end{equation}
Since all the constants appearing in this expression are finite and positive, as well as  $z \approx z_c + \tfrac{x^2}{z_c^2 \omega_-}$ in that region, this means that it can be approached in finite affine time. Nonetheless, the presence of a minimum in the radial function $z(x)$, now interpreted as a {\it bounce}, allows any such geodesic to be uniquely extended and traverse this region in a smooth way, as depicted in Fig. \ref{fig:nullradialEHEiBI}. This behaviour is typically identified as the presence of a wormhole \cite{VisserBook}, with $z=z_c$ is throat. In this plot, we make use of the radial coordinate $x$ to show that effective null geodesics cross the region  $x=0$ ($z=z_c$) in a different (but finite) affine time as compared to their background geodesics counterpart. Conversely, GR geodesics meet instead their incompleteness  there due to the lack of an $x<0$ region (and because the extension at the zero-area surface $x=0$ would not be unique). For non-radial geodesics, $L \neq 0$, and given the finiteness of both functions $H(z)$ and $h(z)$ at every $z \geq z_c$, then Eq.(\ref{eq:uzEiBI}) will have orbits that reach a turning point but also others that can overcome the potential barrier to reach the $x=0$ region in finite affine time. However, and similarly as the radial effective null geodesics, and the radial and non-radial in the background ones, nothing prevents any such geodesic to continue their trip towards the $x<0$ region. One thus concludes the completeness of all null effective geodesics in EiBI gravity coupled to EH electrodynamics.

\section{Conclusion}

In this work we have considered three spherically symmetric geometries found within quadratic $f(R)$ and EiBI gravity in a Palatini formulation with independent metric and affine connection, and studied the completeness of their effective null geodesics caused by the non-linearity of the electromagnetic fields (Born-Infeld and Euler-Heisenberg) souring them. Likewise in systems with linear (Maxwell) fields, the geodesic restoration is achieved either via a displacement of the potentially pathological region to the boundary of the space-time (in the $f(R)$ case), unreachable in finite affine time by any set of geodesics; or by development of a bounce in the radial function (in the EiBI case), interpreted as the throat of a wormhole structure, which allows some of such effective (radial and $L\neq 0$ alike) geodesics to pass through it and freely expand to other regions of the space-time. Within GR, any attempt to implement each of these mechanisms is doomed to introduce unpleasant features in the matter fields or pathologies in the causal structure of the manifold due to the surveillance of the singularity theorems.

In this sense, one should note that while the effective metric arises as a consequence of the non-linearity of the matter sector, what actually {\it cures} the singularity is the non-linearity in the gravitational field; in other words, the gravitational mechanism is robust against modifications in the matter sector, both in the background geometry and in the propagation one. The completeness of these effective null geodesics thus wraps up the core results of our studies on this topic: that singularity-removal of geodesically incomplete GR solutions arises naturally in some Palatini-based geometries, that it is achieved thanks to these two mechanisms, and that it is supported according to (background and effective) geodesic completeness, impact of tidal forces on extended (time-like) observers, scattering of waves, and accelerated observers.

Beyond purely theoretical considerations, the propagation of photons on an effective metric has repercussions in observable settings, most notably in the observational properties of compact objects supported and/or surrounded by a material described by non-linear matter fields. For instance, in order to find their optical appearance when illuminated by an accretion disk one needs to integrate the effective geodesic equation rather than the background one, even within GR \cite{Kruglov:2020tes,Zeng:2022pvb,Wen:2022hkv}, with important observational consequences such as in shadow images size \cite{Vagnozzi:2022moj}. Theories of the kind considered in this work will modify the background metric, the effective geodesic equation, and the gravitational wave generation and propagation  \cite{BeltranJimenez:2017uwv} inside the matter sources while leaving the weak-field limits mostly unaffected \cite{Jana:2017ost}, thus yielding observational signatures that may significantly differ from their GR counterparts  \cite{Olmo:2023lil}. As  such, they represent a suitable theoretical framework allowing to solve one of the most acute problems GR has while at the same time offering solutions that can be put to observational test using the tools of multi-messenger astronomy  \cite{Addazi:2021xuf}.

\section*{Acknowledgements}

Work supported by Grants PID2019-108485GB-I00 and PID2020-116567GB-C21, funded by MCIN/AEI/10.13039/501100011033, and the project PROMETEO/2020/079 (Generalitat Valenciana).  Further support is provided by the EU's Horizon 2020 research and innovation (RISE) programme H2020-MSCA-RISE-2017 (FunFiCO-777740) and  by  the  European Horizon  Europe  staff  exchange  (SE)  programme HORIZON-MSCA-2021-SE-01 (NewFunFiCO-10108625). This article is based upon work from COST Actions CA18108 and CA21136.

\end{document}